\documentclass[prl,aps,preprint,superscriptaddress,showpacs,amsmath,amssymb,floatfix]{revtex4}
                                                                                \usepackage{graphicx}

\newcommand{\be}{\begin{equation}}
\newcommand{\ee}{\end{equation}}
\newcommand{\wz}{\omega_{z}}
\newcommand{\rp}{r_{\perp}}
\newcommand{\pp}{p_{\perp}}

\begin{document}

\title{Umklapp collisions and center of mass oscillation of a
trapped Fermi gas}
\author{G. Orso}
\affiliation{Dipartimento di Fisica, Universita' di Trento and
BEC-INFM, 1-38050 Povo, Italy}

\author{L.P. Pitaevskii}
\affiliation{Dipartimento di Fisica, Universita' di Trento and
BEC-INFM, 1-38050 Povo, Italy}
\affiliation{Kapitza Institute for Physical Problems, 117334 Moscow, Russia}

\author{S. Stringari}
\affiliation{Dipartimento di Fisica, Universita' di Trento and
BEC-INFM, 1-38050 Povo, Italy}

\date{\today}

\begin{abstract}
Starting from the the Boltzmann  equation, we study the center of mass
oscillation of
a harmonically trapped normal Fermi gas in the presence of a
one-dimensional periodic potential. We show that for  values of the the
Fermi energy above the first Bloch band the center of
mass motion is strongly damped in the collisional regime due to umklapp
processes. This should be contrasted with the behaviour of a superfluid
where one instead expects the occurrence of persistent Josephson-like
oscillations.
\end{abstract}
\maketitle

It is well known that the center of mass of an atomic cloud confined in a
harmonic potential oscillates with the trap frequency irrespective of 
temperature, inter-particle interactions and quantum statistics 
(Khon's theorem).
This general result is no longer true if the system is also confined
by a  periodic potential generating barriers separating different wells.
While a Bose-Einstein condensate still exhibits a collective behaviour
since atoms can tunnel in a coherent way through the barriers \cite{lens},
a non interacting Fermi gas with Fermi energy lying above the first Bloch
band is not able to oscillate around the minimum of the harmonic well
\cite{luca}. The sudden shift of the harmonic trap indeed causes an
asymmetry in the occupation of the left and right orbits which, due to
their open nature, persists in time. This  causes the system to remain
trapped at one side of the harmonic field.  The inclusion of interactions
  changes this scenario favouring the relaxation  towards
equilibrium \cite{lens2}.

An important  question is to understand what is the behaviour of a normal Fermi
gas in the collisional regime where the system is in  conditions of
local equilibrium and  the dynamic behaviour  is described by 
the equations of hydrodynamics. In
particular the main question addressed in the present letter is whether in
the collisional regime  the system is able to exhibit center of mass
oscillations. The problem is relevant in view
of the search of signatures of superfluidity in ultra cold Fermi gases.

We consider a two-component atomic gas 
trapped by an external potential given by the sum of 
a harmonic trap of magnetic origin and of a stationary
 optical potential modulated along the $z$-axis. 
 The resulting potential is given by 
  \begin{equation}
 V_{ext} = {1\over 2} m \left(\omega_{\perp}^2 \rp^2 + \omega_z^2z^2
 \right) + s E_R\sin^2q_Bz, 
 \label{V}
 \end{equation}
 where $\omega_{\perp}$, $\omega_z $ are the frequencies 
of the harmonic  
trap, $E_R  = \hbar^2q_B^2/2m$ is the recoil energy, $q_B$ being the 
Bragg momentum and $s$ is a 
 dimensionless parameter providing  the intensity of the 
 laser beam.  
 The optical potential has periodicity 
 $d=\pi /q_B$  along the $z$-axis. 
In the semiclassical approximation, the energy of 
the resulting Bloch states is given by 
$\epsilon(\mathbf r,\mathbf p)=\epsilon(p_z)+\pp^2/2m+ V_{ho}(\mathbf r)$, 
where $V_{\rm ho}=m(\omega_\perp ^2 r_\perp ^2+\wz^2z^2)/2$ and $\epsilon(p_z)$ is the dispersion relation of the lowest band
obtained by solving the 1D Schrodinger equation with the Hamiltonian
$H=p_z^2/2m+sE_R\sin^2(q_B z)$. Since $\epsilon(p_z)$ is a 
periodic function with period $2\hbar q_B$, the quasi-momentum
$p_z$ is restricted to the first Brillouin zone defined by 
$-\hbar q_B \leq p_z \leq \hbar q_B$.
We restrict our discussion
to low temperatures $T \ll T_F$ and neglect higher bands.  
We consider situations  where the 
system is in the collisional (hydrodynamic) regime
and we calculate the relaxation rate of the center of mass oscillation due to
umklapp collisions. In order to achieve the hydrodynamic condition at such low
temperatures, where collisions are suppressed by the Pauli principle, one
should increase the value of the scattering length working close to a
Feshbach resonance and/or work with very shallow traps along the $z$-th
direction.

We begin our analysis by defining the center of mass coordinates and momenta
as 
$Z=\int z fd\mathbf p d\mathbf r/h^3$,
$P_z=\int p_{z}f d\mathbf p d\mathbf r/h^3$ 
where $f=f_\uparrow+f_\downarrow$ and $f_\uparrow,f_\downarrow$
are the distributions functions of the two spin species which 
are assumed to be equal ($f_\uparrow=f_\downarrow=f/2$). 

 By suitable integrations of the
Boltzmann equation, one finds the following {\sl exact} equations
for the center of mass oscillation:
\begin{eqnarray}
\label{cont}
\frac{\partial}{\partial t}Z-\int \frac{\partial \epsilon}{\partial p_z}f 
\frac{d{\mathbf p}d{\mathbf r}}{h^3}=0
\,,
\\
\frac{\partial}{\partial t}P_z+m\wz^2Z=\int p_z C
\frac{d{\mathbf p}d{\mathbf r}}{h^3}
\,,
\label{umk}
\end{eqnarray}
where  $C$ is the collisional integral describing $s-$wave
scattering between Bloch states. Eq.(\ref{cont}) directly
follows from the equation of continuity, $j_z=\partial\epsilon /\partial p_z$ 
being the  current density along $z$. 
In the absence of the periodic potential, the integral in Eq.(\ref{umk})
is zero as momentum is rigorously conserved during collisions.
 Since in this case the dispersion law reduces to the 
free value $p_z^2/2m$, one has 
$\int \partial \epsilon /\partial p_z fd\mathbf p d\mathbf r/h^3=P_z/m$
and one recovers the general result $\omega=\wz$ for the 
frequency of the oscillation.
 
The effect of the optical lattice on the above equations is twofold. 
First it changes the dispersion relation from the free value $p_z^2/2m$
to $\epsilon(p_z)$. 
Second, the traslational symmetry is broken and momentum 
is no longer a good quantum number \cite{peierls}. 
This has dramatic consequences 
for the collisional integral as momentum conservation 
is replaced by the weaker constraint 
$p_{1z}+p_{2z}-p_{3z}-p_{4z}=2 \hbar n q_B$,
where $n$ is an integer and
$p_{1z},p_{2z}$ and $p_{3z},p_{4z}$ are, respectively, 
the initial and final quasi-momenta of the two colliding 
particles.
For two-body interactions, one finds that only
umklapp processes with $n=\pm 1$ are allowed.
  
In an umklapp collision, the system exchanges momentum $\pm 2\hbar q_B$
with the optical lattice. This means that $P_z$
varies in time not only because of the oscillator force 
$F=-m\wz^2z$ but also as
a result of umklapp processes, which 
act to damp the oscillation.
We therefore expect that the right hand side of Eq.(\ref{umk})  
is non-zero for out-of-equilibrium distributions.

Let us write the collisional integral  
as $C(\mathbf p,\mathbf r)=C^+(\mathbf p,\mathbf r) - C^-(\mathbf p,\mathbf r)$, 
where $C^+(\mathbf p,\mathbf r)$ and $C^-(\mathbf p,\mathbf r)$ describes collisions in which 
one of the particles has, respectively, final and initial 
momentum  $\mathbf p$.
Introducing the compact notation 
$d\mathbf p \equiv d\mathbf p_\perp \Theta (|\hbar q_B|-p_z)dp_z$, 
where $\Theta(x)$ is the step function,
the $C^+$ term  can be conveniently 
expressed via the Fermi golden rule in the general form
$C^+(\mathbf p_1,\mathbf r)=\int d\mathbf p_{2} 
d\mathbf p_{3} d\mathbf p_{4}
D$, with
\begin{eqnarray}
\label{coll}
&&D(\mathbf p_{1},\mathbf p_{2},\mathbf p_{3},\mathbf p_{4},\mathbf r)
=\frac{2}{h^6}  
\frac{2\pi}{\hbar}\nonumber\\   
&&\sum_{n=0,\pm 1}|U_{fi}^{(n)}|^2 \delta (p_{1z}+p_{2z}-p_{3z}-p_{4z}-2n q_B)\nonumber\\
&&\delta \big( \mathbf p_{\perp 1}+\mathbf p_{\perp 2}-
\mathbf p_{\perp 3}-\mathbf p_{\perp 4} \big)
 \delta \big(\epsilon(1)+\epsilon(2)-\epsilon(3)-\epsilon(4)\big)\nonumber\\
&&  f_\uparrow(3)f_\downarrow(4)
\big(1-f_\uparrow(1)\big) \big(1-f_\downarrow(2)\big),
\end{eqnarray}
where $U_{fi}^{(n)}$ is the $n-$dependent matrix element of
the two-body interaction and
$\epsilon(j)=\epsilon(\mathbf r,\mathbf p_j)$. 
The factor $2$ in Eq.(\ref{coll})  comes
from the trace over spin indices and $f_\sigma$ are normalized to 
$\int f_\sigma d\mathbf p d\mathbf r/h^3=N/2$.
The $C^-$ term is simply obtained from $C^+$ by interchanging  
initial and final states. 
At thermal equilibrium one has 
$f_\sigma=f_0=[e^{\beta (E-\mu)}+1]^{-1}$ with 
$\beta=1/k_BT$ and $E=\epsilon(\mathbf r,\mathbf p)$. In this case
$C^+=C^-$ and hence $C=0$.
The total number of collisions per unit time is given by
\be\label{gamma}
\Gamma=\int C^+ \frac{d\mathbf p_1
d\mathbf r}{h^3}=\Gamma_{nor}+\Gamma_{uk},
\ee
where we have written explicitly the contributions 
$\Gamma_{nor}$ and $\Gamma_{uk}$  coming from normal $(n=0)$
and umklapp $(n=\pm 1)$ collisions [see Eq.(\ref{coll})].  

While in the general case Eqs. (\ref{cont}) and (\ref{umk})
are not sufficient to calculate the frequency of the mode, in the 
hydrodynamic regime one can make the ansatz 
 \be\label{ansatz}
f_\sigma ({\mathbf r,\mathbf p},t)=
f_0(\epsilon({\mathbf r,\mathbf p})+u(t)p_z),
\ee
where $u(t)$ is a time dependent function. This corresponds to a 
rigid displacement of the density current $j_z$: 
$j_z (p_z)\longrightarrow j_z(p_z)+u$ and permits to write Eqs (\ref{cont})
and (\ref{umk}) 
in a closed form.

The ansatz (\ref{ansatz}) is valid in the limit of strong collisions
$\omega \tau \ll 1$, where $\omega$ is the frequency of the 
collective oscillation and  $\tau$ is a typical collisional time.
In the classical regime one has $\tau^{-1} \sim \Gamma/N$.
At low temperatures $T \ll T_F$, only the particles near the Fermi surface
can be scattered, because of Fermi statistics, and
one finds that the hydrodynamic condition takes the form 
\be\label{condition}
\frac{1}{\tau} \sim \frac{E_F}{k_B T}\frac{\Gamma}{N} \gg \omega.
\ee
Notice that at low $T$ the rate $\Gamma$ behaves like 
${(T/T_F)}^3$~\cite{vichi}
so that $1/\tau \sim {(T/T_F)}^2$ coincides, apart from a numerical
factor of the order of unity, with the quasiparticle lifetime 
calculated at the Fermi surface \cite{pines}.

Since both the integrated current $\int j_z dz$ and the collisional
integral are zero at equilibrium, we
expand the ansatz (\ref{ansatz}) to first order in $u$ 
and plug the result into
Eqs.(\ref{cont}) and (\ref{umk}). 
Analogously, for the center-of-mass momentum one finds
 $P_z=\int p_z fd\mathbf p d\mathbf r/h^3=u \tilde{m}N$,
where 
\be
\label{mass}
\tilde{m}=-\frac{2}{N}\int p_z^2 
\frac{\partial f_0 }{\partial \epsilon(p_z)}\frac{d\mathbf p d\mathbf r}{h^3}
\ee
plays the role of an effective mass for the 
center of mass oscillation \cite{comment}. 
This permits us to cast 
Eqs.(\ref{cont}) and (\ref{umk}) in the following closed form, holding
in the limit of small amplitude oscillations:
\begin{eqnarray}
\label{ma}
\frac{\partial}{\partial t}Z=\frac{P_z}{\tilde{m}}\,,
\\
\frac{\partial}{\partial t}P_z+m\wz^2Z=-\frac{P_z}{\tau_{uk}}\,,
\label{ukrate}
\end{eqnarray}
where 
\be\label{umklapp}
\frac{1}{\tau_{uk}}=-\frac{1}{k_B T}\frac{\int p_{1z}(p_{1z}+p_{2z}-p_{3z}-p_{4z}) 
D d{\mathbf p}_1d{\mathbf r}}{  \int p_z^2 \partial f_0/
\partial \epsilon(p_z) d{\mathbf p}_1 d{\mathbf r}}
\ee
defines the relevant relaxation time of the oscillation
due to umklapp collisions.
To derive Eq.(\ref{umklapp}) we have used the 
identity $
\sum_{j=1}^4 p_j \partial D /\partial \epsilon(p_{jz})=
-\beta (p_{1z}+p_{2z}-p_{3z}-p_{4z})D$,
holding at equilibrium also for Fermi statistics.
Notice that the contribution from the 
normal $(n=0)$ collisions, which conserve momentum,  
identically vanishes. 
Equations (\ref{ma}) and (~\ref{ukrate}) have the form of a damped 
harmonic oscillator.    
Looking for solutions of the form $e^{-i\omega t}$ the dispersion law 
is given by 
\be
\omega=-\frac{i}{2 \tau_{uk}}\pm \sqrt{\frac{m}{\tilde{m}}\wz^2-\frac{1}
{{(2\tau_{uk})}^2}}
\ee
showing that the oscillations become overdamped if 
$\wz \tau_{uk}<\frac{1}{2}(\frac{\tilde{m}}{m})^{1/2} $.

By recalling that the integrand $D$ in Eq.(\ref{coll}) is symmetric under
interchange  $1 \rightleftharpoons 2$ and antysimmetric under 
interchanges  $1\rightleftharpoons 3$ or $1\rightleftharpoons 4$,
 the factor $p_{1z}$ in  
the numerator can be substituted by the combination 
$(p_{1z}+p_{2z}-p_{3z}-p_{4z})/4$ which is equal to 
$\pm \hbar q_B/2$ for $(n=\pm 1)$ umklapp processes. 
Taking into account Eq.(\ref{mass}),  
Eq.(\ref{umklapp}) can then be written as 
\be\label{uk}
\frac{1}{\tau_{uk}}= 4\frac{m}{\tilde{m}}\frac{E_R}{k_BT}\frac{\Gamma_{uk}}{N},
\ee
where $\Gamma_{uk}$ is defined by Eqs.(\ref{coll}) and (\ref{gamma}). 

Our next goal is to show that, for sufficiently tight optical lattices and
gas densities corresponding to $T_F/2\delta >1$, 
the condition (\ref{condition}) automatically implies the overdamping of the 
center of mass oscillation. 
In the following we  consider relatively large
laser intensities and work in tight-binding approximation.
Under this assumption, the energy dispersion is given by 
$\epsilon(p_z)=\delta \big( 1-\cos[\frac{p_z d}{\hbar}]\big)$ where 
the bandwidth $2\delta$ is proportional to the tunneling rate
between consecutive wells.
The Fermi energy is related to the total number of particles
by $N=2 \int \Theta(T_F-\epsilon(\mathbf p,\mathbf r))
d\mathbf pd \mathbf r/h^3$, yielding
\be\label{number}
N=\frac{16}{15 \pi^2}\Big(\frac{E_R}{\delta}\Big)^{1/2}
\frac{\delta^{3}}
{\hbar^3 \omega_\perp^2 \wz} \int_{-\pi}^{\pi}
h(\tilde{p})^{5/2} \Theta(h(\tilde{p}))d\tilde{p},
\ee
where $\tilde{p}=p_z d/\hbar$ and  $h(\tilde{p})=T_F/\delta-1+\cos \tilde{p}$.
Eq.(\ref{number}) permits to calculate $T_F$ as a function of the
free Fermi energy $T^0_F={(3N)}^{1/3}\hbar ({\omega_\perp}^2\wz)^{1/3}$, 
evaluated in the absence of the optical lattice, the recoil energy $E_R$ 
and the bandwidth $2 \delta$.
Neglecting higher bands, the Fermi energy $T_F$ is always 
smaller than the free value $T^0_F$ 
reflecting the fact that $\epsilon(p_z) \leq p_z^2/2m$ in the
first Brillouin zone. 

In order to calculate the relevant interaction matrix elements $U_{fi}$,
we first neglect harmonic trapping.  
The scattering states   
are $e^{i \mathbf\pp \cdot \mathbf \rp}\psi_k(z)$, 
where $\psi_k(z)$ 
is the 1D Bloch wavefunction with quasi-momentum $k$. 
We make the  
ansatz  $\psi_k(z)\sim \sum_l e^{i p k d/\hbar}f(z-ld)$, 
where $l$ labels the wells, $f$ is localized 
at the origin and normalized to $\int_{d/2}^{d/2}|f(z)|^2=1$.
In the following we will consider  
a delta function potential $U(r)=g\delta(r)$, the coupling constant 
$g$ being related to
the scattering length $a$ by $g=4\pi \hbar^2 a/m$.
In the tight-binding limit, the matrix elements 
are given by
\be\label{matrix}
U_{fi}^{(\pm 1)}=U_{fi}^{(0)}=g d\int_{-d/2}^{d/2}f^4(z)dz=g \alpha.
\ee
Conversely, in the absence of the optical lattice the eigenstates of $H$
are simply plane waves and the matrix elements are given by
$U_{fi}^{(\pm 1)}=0$ and $U_{fi}^{(0)}=g$.
The factor $\alpha$ of Eq.(\ref{matrix})
is larger than one and increases by increasing the laser 
intensity $s$, the wavefunction
$f(z)$ becoming more and more peaked at the origin.
For very large values of $s$, one can use the asymptotic 
result $\alpha=\sqrt{\pi/2}s^{1/4}$.
This formula actually overestimates the correct value of 
$\alpha$ for smaller $s$. As an
example, for $s=5$ it gives $\alpha=1.9$ while
a more accurate calculation gives $\alpha=1.6$ \cite{pedri}. 
 From the above discussion, we conclude that
 the periodic potential enhances interaction 
effects in two different ways. First it allows for additional 
(umklapp) collisions to take place. Second, 
the optical confinement compresses the gas and this results in an
increase of the coupling constant from  $g$ to $\alpha g$.

\begin{figure}
\begin{center}
\includegraphics[height=8cm,angle=270]{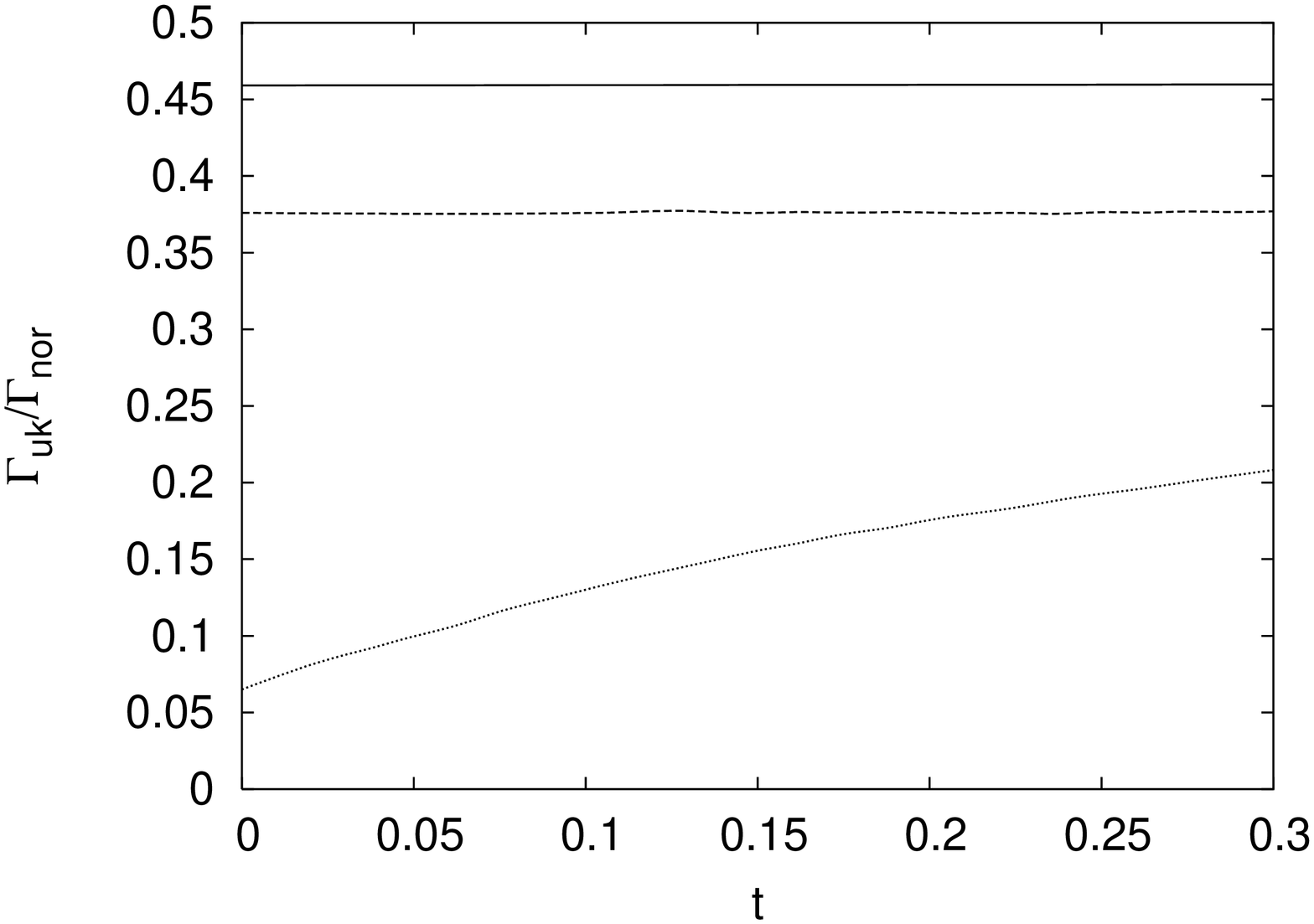}
\caption{Ratio of umklapp over normal collisional rates [see Eq.(\ref{gamma})]
as a function of $t=T/T_F$ for the values
$T_F/2\delta=5$ {\sl (solid line)}, $2$ {\sl (dashed line)}, $1$ {\sl (dotted line)}.}
\label{fig1}
\end{center}
\end{figure}

In the presence of harmonic trapping, we can still use result (\ref{matrix})
provided local density approximation is applicable.
This requires that the trapping frequencies should be small
compared to the Fermi energy and the bandwith.
The collisional rates $\Gamma_{nor}$ and $\Gamma_{uk}$
appearing in Eqs.(\ref{coll}) and (\ref{gamma}) have been integrated 
using standard Montecarlo techniques.
The ratio $\Gamma_{uk}/\Gamma_{nor}$ is  plotted in Fig.\ref{fig1} 
as a function of the temperature
for different values of the parameter $T_F/2\delta$. 
For $T_F \ll 2\delta$, umklapp collisions 
are negligible at low temperatures. In fact, in this case,
the typical initial quasi-momenta $p_{1z},p_{2z}$
are small compared to $\hbar q_B$ and, due
to the energy conservation, this is also true for the 
final quasimomenta, meaning that processes with $n \neq 0$
are unlike.
When the Fermi energy 
is larger than the bandwith, umklapp processes become instead
competitive with the normal ones even at low temperatures
and we see that
the ratio $\Gamma_{uk}/\Gamma_{nor}$ saturates to a constant value.
In a uniform system this constant can be analytically
evaluated and  is equal to $1/2$. In the trapped case
 one finds a smaller value because
the effective Fermi energy $T_F(r)=T_F-V_{ho}(r)$
is $r-$dependent and therefore umklapp collisions are quenched 
at the edge of the cloud where $T_F(r) <2\delta$.

\begin{figure}
\begin{center}
\includegraphics[height=8cm,angle=270]{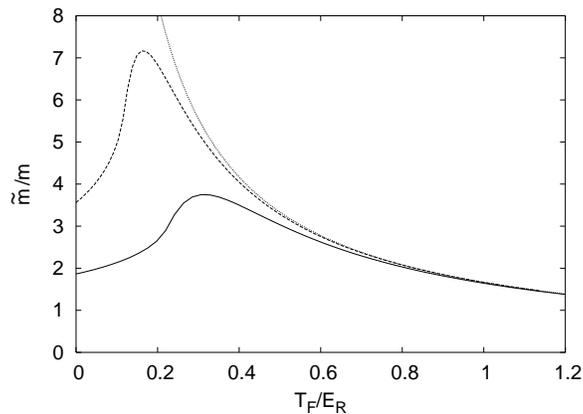}
\caption{Effective mass of the dipole oscillation [see Eq.(\ref{mass})]
plotted at $T=0$ as a function of the parameter $T_F/E_R$ for laser 
intensities $s=5$ {\sl (solid line)} and 
$s=8$ {\sl (dashed line)}. The asymptotic curve $5E_R/3T_F$ is also shown 
{\sl (dotted line)}.}
\label{fig-massa}
\end{center}
\end{figure}

We are now ready to compare the overdamping condition
 with the hydrodynamic condition (\ref{condition}). 
To this purpose, we have evaluated the effective mass $\tilde{m}$ for 
the center of mass oscillation introduced in Eq.(\ref{mass}). 
At $T=0$ and for a fixed laser 
intensity $s$, the ratio 
$\tilde{m}/m$ is a function of the parameter $T_F/E_R$.
This function is plotted in Fig.(\ref{fig-massa}) for different values
of $s$. The figure shows that, for $T_F/E_R \gtrsim 1$, 
the ratio $\tilde{m}/m$ does not depends on the laser  
intensity and is of the order of unity. 

In the following we will consider typical configurations
with $T_F \sim E_R$.
By comparing 
Eq.(\ref{condition}) with Eq.(\ref{uk}), we conclude that, due 
to umklapp processes, 
the center of mass oscillation of a trapped gas
confined by a tight optical potential with $T_F>2\delta$ will be 
overdamped in the collisional regime $\wz \tau \ll 1$ since in
this case $\Gamma_{uk}\sim\Gamma_{nor},\;\tilde{m}\sim m $ and therefore
$\tau_{uk}\sim \tau \ll \wz^{-1}$.

\begin{figure}
\begin{center}
\includegraphics[height=8cm,angle=270]{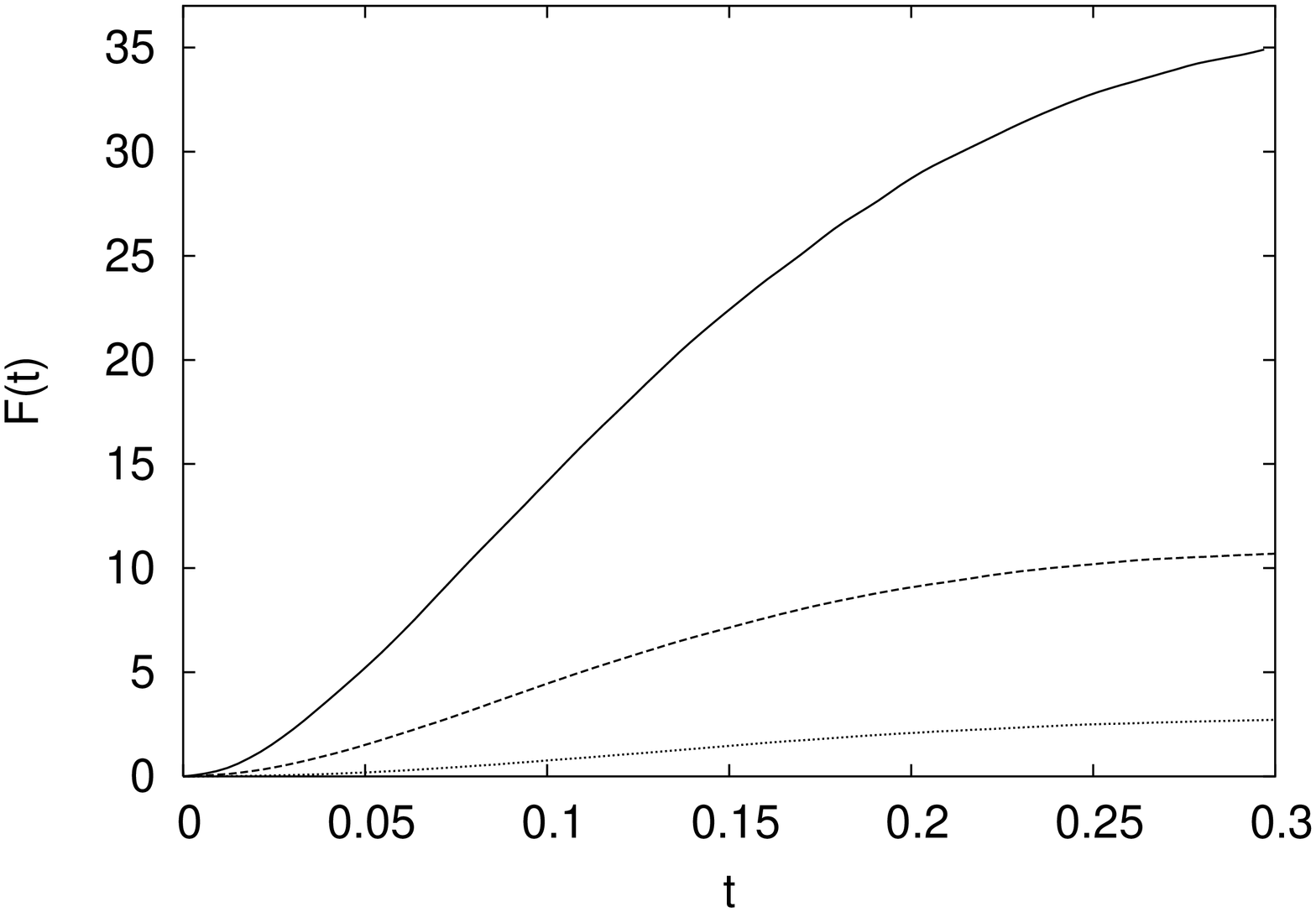}
\caption{Dipole relaxation function $F$ 
[see Eq.(\ref{final})] as
a function of the reduced temperature $t=T/T_F$ for the values
$T_F/2\delta=5$ {\sl (solid line)}, $2$ {\sl (dashed line)}, $1$ {\sl (dotted line)}.}
\label{fig2}
\end{center}
\end{figure}

The relaxation rate of the dipole oscillation can be conveniently
written in the form:
\be
\label{final}
\frac{1}{\tau}_{uk}=\alpha^2 \frac{a^2}{d^2}\frac{\delta}{\hbar}F\Big(\frac{T}{T_F},\frac{T_F}{2\delta}\Big)
\ee
where the temperature dependence of the function 
 $F$ is plotted in Fig.\ref{fig2} for different values of
the paramter $T_F/2\delta$. 
The possibility for the system to be in the overdamping 
regime depends on the actual values of the parameters in Eq.(\ref{final}).
As a concrete example, let us consider a two-component gas of $N=10^5$ 
potassium $(^{40}K)$ 
atoms with trap frequencies $\omega_\perp=2 \pi \cdot 275$Hz and
$\wz=2 \pi \cdot 24 $Hz, corresponding to $T^0_F= 390nK$.
For the optical lattice we assume 
$s=5$ and periodicity $d=400$nm, corresponding to  
$E_R=9.2 \delta=374 nK$.
For the Fermi energy one then
finds $T_F = 0.85 T_F^0$ and from the value of $T_F/2\delta$
one finds $\tilde{m}/m=1.8$. 
We see from Fig.\ref{fig2} that the overdamping condition 
$\wz \tau_{uk} \ll 1$ is 
satisfied even at low temperatures $T/T_F\sim 0.05-0.1$,
provided the scattering
length is moderately large, say $|a|/d \geq 0.1-0.2$.
This can be accomplished 
experimentally working close to a Feshbach resonance. 

For lower temperatures or smaller scattering lengths, the system 
enters the collisionless regime. Since in this regime
 the center-of-mass oscillation is self-trapped 
under the same $T_F>2\delta$ condition \cite{luca}, 
we conclude that the system can never exhibit 
undamped center-of-mass oscillation in the normal phase.
In the superfluid phase, on the contrary, one expects the occurrence of
persistent (undamped) Josephson-like 
oscillations \cite{wouters}. In particular, in the weak-coupling (BCS) limit,
where the distribution function $f$ does not
deviate significantly from the ideal gas value,
the frequency of the dipole oscillation is expected to be 
$\wz \sqrt{m/\tilde{m}}$
with $\tilde{m}$ given by Eq.(\ref{mass}) and Fig.\ref{fig-massa}. 

Let us finally discuss the conditions of applicability of our approach.
These concern the stability of the Bloch states
with respect to the interaction.
First, the scattering length must be small compared to the
interwell distance, i.e. $a \ll d$, otherwise  
we cannot model the interaction with a $\delta$-function 
potential.
Second, the broadening of the Bloch wavefunction due to collisions  
should be smaller than the bandwith: $\hbar/\tau \ll \delta$.
This condition is needed in order to apply the semiclassical
Boltzmann picture and can be reasonably well satisfied
for the temperatures of interest. 

This work was supported by the Ministero dell'Istruzione, 
dell'Universita' e della Ricerca (M.I.U.R.).

\newpage


\begin{thebibliography}{9}
\bibitem{comment} This mass should not be confused with the
effective mass $m^*=2 E_R/ \pi^2 \delta $ calculated at the bottom 
of the band.
At $T=0$, one finds $\tilde{m}=m^*$ if $T_F \ll 2 \delta$.
\end{thebibliography}
\end{document}